\documentstyle[12pt,epsf,enumerate]{article}

\begin{document}
\def\comment#1{\marginpar{{\scriptsize #1}}}
\def\framew#1{\fbox{#1}}                        
\def\framep#1{\noindent \fbox{\vbox{#1}}}       
\def\framef#1{\fbox{\vbox{ #1 }}}               
\def\goto{$\ra $ }

\def\be{\begin{equation}}
\def\ee{\end{equation}}
\def\bea{\begin{eqnarray}}
\def\eea{\end{eqnarray}}

\newtheorem{proposition}{Proposition}[section]
\def\bprop{\bigskip\begin{proposition}~~~\\ \rm}
\def\eprop{\end{proposition}\bigskip}
\def\proof{\bigskip \noindent {\it Proof.} \ }
\newtheorem{naming}{Definition}[section]   
\def\bnam{\bigskip\begin{naming}~~~\\ \rm}
\def\enam{end{naming}\bigskip}

\bibliographystyle{unsrt}
\def\br{\begin{thebibliography}{abc}}
\def\er{\end{thebibliography}}
\def\rf{\bibitem}
\def\cstars{$C^*$-algebras }
\def\cstar{$C^*$-algebra }
\def\unit{I\!\!I}
\def\norm#1{\parallel #1 \parallel}
\def\abs#1{\left| #1\right|}
\def\ha{\widehat{\cal A}}
\def\hc{\widehat{\cal C}}
\def\prim{$Prim \ca~$}

\def\bar#1{\overline{#1}}
\def\what{\widehat}
\def\wtilde{\wtilde}
\def\sp{~~~~~}
\def\bra#1{\left\langle #1\right|}
\def\ket#1{\left| #1\right\rangle}
\def\EV#1#2{\left\langle #1\vert #2\right\rangle}
\def\VEV#1{\left\langle #1\right\rangle}
\def\pa{\partial}
\def\del{\nabla}

\def\a{\alpha}
\def\b{\beta}
\def\c{\raisebox{.4ex}{$\chi$}}
\def\d{\delta}
\def\e{\epsilon}
\def\f{\phi}
\def\g{\gamma}
\def\h{\eta}
\def\i{\iota}
\def\j{\psi}
\def\k{\kappa}
\def\l{\lambda}
\def\m{\mu}
\def\n{\nu}
\def\o{\omega}
\def\p{\pi}
\def\q{\theta}
\def\r{\rho}
\def\s{\sigma}
\def\t{\tau}
\def\u{\upsilon}
\def\x{\xi}
\def\z{\zeta}

\def\D{\Delta}
\def\F{\Phi}
\def\G{\Gamma}
\def\J{\Psi}
\def\L{\Lambda}
\def\O{\Omega}
\def\P{\Pi}
\def\Q{\Theta}
\def\S{\Sigma}
\def\U{\Upsilon}
\def\X{\Xi}
\def\Z{\Zeta}

\def\ca{{\cal A}}
\def\cb{{\cal B}}
\def\cc{{\cal C}}
\def\cd{{\cal D}}
\def\ce{{\cal E}}
\def\cf{{\cal F}}
\def\cg{{\cal G}}
\def\ch{{\cal H}}
\def\ci{{\cal I}}
\def\cj{{\cal J}}
\def\ck{{\cal K}}
\def\cl{{\cal L}}
\def\cm{{\cal M}}
\def\cn{{\cal N}}
\def\co{{\cal O}}
\def\cp{{\cal P}}
\def\cq{{\cal Q}}
\def\czr{{\cal R}}
\def\cs{{\cal S}}
\def\ct{{\cal T}}
\def\cu{{\cal U}}
\def\cv{{\cal V}}
\def\cw{{\cal W}}
\def\cx{{\cal X}}
\def\cy{{\cal }}
\def\cz{{\cal Z}}
%
%
\def\inbar{\,\vrule height1.5ex width.4pt depth0pt}
\def\IG{\relax\,\hbox{$\inbar\kern-.3em{\rm G}$}}
\def\IU{\relax\,\hbox{$\inbar\kern-.3em{\rm U}$}}
\def\ID{\relax{\rm I\kern-.18em D}}
\def\IF{\relax{\rm I\kern-.18em F}}
\def\IH{\relax{\rm I\kern-.18em H}}
\def\II{\relax{\rm I\kern-.17em I}}
\def\I1{\relax{\rm 1\kern-.28em l}}
\def\IM{\relax{\rm I\kern-.18em M}}
\def\IN{\relax{\rm I\kern-.18em N}}
\def\IP{\relax{\rm I\kern-.18em P}}
\def\IQ{\relax\,\hbox{$\inbar\kern-.3em{\rm Q}$}}

\def\IC{\hbox{{\bf \inbar}\hskip-4.0pt C}}

\def\IR{\hbox{I\hskip-1.7pt R}}
\font\cmss=cmss10 \font\cmsss=cmss10 at 7pt
\def\IZ{\relax\ifmmode\mathchoice
{\hbox{\cmss Z\kern-.4em Z}}{\hbox{\cmss Z\kern-.4em Z}}
{\lower.9pt\hbox{\cmsss Z\kern-.4em Z}}
{\lower1.2pt\hbox{\cmsss Z\kern-.4em Z}}\else{\cmss Z\kern-.4emZ}\fi}
%
\def\Up{\Uparrow}
\def\up{\uparrow}
\def\Dn{\Downarrow}
\def\dn{\downarrow}
\def\Rt{\Rightarrow}
\def\rt{\rightarrow}
\def\Lt{\Leftarrow}
\def\lt{\leftarrow}
\def\bc{{\bf{C}}}

\newcommand{\paf}[2]{\frac{\partial#1}{\partial#2}}
\renewcommand{\thefootnote}{\fnsymbol{footnote}}
\def\fn{\footnote}
\footskip 1.0cm

\thispagestyle{empty}
\setcounter{page}{0}

\hfill \today

\vskip2cm

\centerline {\Large {\bf Mixed Phases for the $t$-$J$ Model}}
\vspace{0.75cm}
\centerline {Elisa Ercolessi$^{a)}$,
             Giuseppe Morandi$^{a)}$,
             Leonardo Pisani$^{b)}$, 
             Marco Roncaglia$^{a)}$ }
\vspace{1cm}
\centerline {\it $^{a)}$Dipartimento di Fisica, Universit\`a di
Bologna, }
\centerline {INFM and INFN, Unit\`a di Bologna, Italy.}
\centerline {\it Via Irnerio 46, I-40126, Bologna, Italy.}
\centerline {\it $^{b)}$Dipartimento di Fisica, Universit\`a di
Camerino}
\centerline {\it Via Madonna delle Carceri, Camerino, Italy.}
\vspace{.5cm}

\begin{abstract}
We study the competition between non-magnetic (dimer or flux) states
and short-range antiferromagnetically ordered RVB states in the
$t$-$J$ model and present a finite temperature phase diagram. 
We show that, for a wide range of temperatures and dopings, the stable
phase results to be a state in which both the flux and the RVB 
parameters are nonzero.

\end{abstract}

\newpage

\setcounter{footnote}{0}

In this paper we want to extend the analysis of the phase
diagram  of the $t$-$J$ model on a square lattice, started by some of the
authors in ref. \cite{EPR}, where only the so called non-magnetic states
(such as the dimer and the flux phase) were considered.

Our starting point is the t-J Hamiltonian, as originally proposed in
\cite{BZA}:
\be
\ch = t\d \sum_{(ij)} \sum_{\a} c^\dagger_{i\a} c_{j\a}
- \frac{J}{4} \sum_{(ij)} \sum_{\a\b} c^\dagger_{i\a} c_{j\a}
c^\dagger_{j\b} c_{i\b} \; , \label{tj}
\ee
where $(ij)$ stands for a sum over ordered nearest-neighbor sites.
This Hamiltonian incorporates the below-half-filling constraint only
in average: the hopping coefficient $t$ of the first (kinetic) term
gets renormalized by the doping factor $\d$
{\footnote[4]{For a detailed discussion of how to derive this
formula in the strong-coupling limit of the Hubbard model at or below
half-filling see \cite{EPR} and references therein.}}. At half filling
($\d=0$) the Hamiltonian reduces to the well-known Heisenberg Hamiltonian, 
with the spin operators written in a fermion representation.
 
As candidates for the ground state of the t-J model, 
the RVB and the flux states have been mainly considered in literature.  
The former, first proposed by Anderson and coworkers \cite{BZA}, is 
characterized by a BCS-type order parameter of the kind 
$\D_{ij} = \langle c_{i\uparrow} c_{j\downarrow} - c_{i\downarrow} 
c_{j\uparrow} \rangle $, which gives short-range
antiferromagnetic correlations. The latter \cite{AM} corresponds to a
nonmagnetic complex order parameter $\chi_{ij} = \langle  \sum_{\a = \uparrow,
\downarrow} c_{i\a}^\dagger c_{j\a} \rangle $, whose phase gives rise to a
nonzero magnetic flux threading the elementary (square) plaquettes of
the lattice in a staggered way. 
It is well known \cite{AZHA} that the Heisenberg Hamiltonian,
and hence the $t$-$J$ model (\ref{tj}) at half filling, is
invariant under local $SU(2)$ transformations. One can exploit
this symmetry \cite{K} to show that the so called mixed-RVB and the
flux phases are degenerate since they are related via one such gauge
transformation. Thus, these apparently different phases actually represent
the same physical state.
Away from half filling, this $SU(2)$-gauge symmetry gets
explicitely broken by the kinetic term. This means that the RVB and the
flux phases are no longer equivalent so that the question of a possible
competition between these two  phases arises.

To start with, we want to reformulate the theory in a way
that enables us to keep always track of the (broken) gauge symmetry.
This can be done by introducing the $2\times 2$ matrix operators:
\be
\Psi_i= \left( \begin{array}{cc} c_{i\uparrow} &  c_{i\downarrow} \\
c_{i\downarrow}^\dagger & - c_{i\uparrow}^\dagger \end{array}
\right)  \label{op}
\ee
and by defining:
\be
\Phi_{ij} = \Psi_i \Psi_j^\dagger\;\; \ \label{p}
\ee
Then, with some algebra, one can easily show that (\ref{tj}) can be
rewritten as:
\be
\ch = t \d \sum_{(ij)} tr[\s_3 \Phi_{ij}] - \frac{J}{8}
\sum_{(ij)} tr[\Phi_{ij}^\dagger \Phi_{ij}] \label{su2}  \;\; ,
\ee
where $\s_3$ is the third Pauli matrix and $tr$ stands for the trace over
$2\times 2$ matrices. It is immediate
to check that the second term in the Hamiltonian is 
gauge invariant under the following $SU(2)$-action:
\be
\Phi_{ij} \longrightarrow g_i  \Phi_{ij} g_j \; \; , \label{suact}
\ee
$g_i$ being any  element of  $SU(2)$ possibly site-dependent, while
the hopping term explicitly  breaks such symmetry. Both the RVB and the
nonmagnetic phases of the $t$-$J$ can be studied in this context. Indeed, 
in the saddle point approximation, the mean value of the fields $\Phi_{ij}$'s
is a function of $\D_{ij}$ and $\chi_{ij}$, namely:
\be
\langle \Phi_{ij}\rangle \, 
\propto \left( \begin{array}{cc} -\chi_{ij}^*  & \D_{ij}  \\
\D_{ij}^* & \chi_{ij} \end{array}
\right) \;\; . \label{mean}
\ee
where $*$ denotes complex conjugation. 
Bearing this in mind, we study the $t$-$J$ Hamiltonian by representing the
partition function in the grand-canonical ensemble by means of a functional
integral over Grasmann fields  representing the fermionic operators
and by decoupling the four-fermion Heisenberg
interaction via a Hubbard-Stratonovich transformation over the auxiliary
$SU(2)$-fields
\be
U_{ij} = \left( \begin{array}{cc} -\chi_{ij}^*  & \D_{ij}  \\
\D_{ij}^* & \chi_{ij} \end{array}
\right) \;\; , \label{aux}
\ee
Following \cite{BZA} and \cite{AM}
we will assume invariance under translations along the lattice for
the parameters $\D$'s and along the diagonal for the $\chi$'s. We are then
left with a total of six independent parameters, $\D_{x,y}$ and
$\chi_{1,2,3,4}$, as shown in Fig. \ref{Fig1}. 
\begin{figure}[t]
\begin{center}
\epsfxsize=7cm

 \epsffile[73 227 531 435]{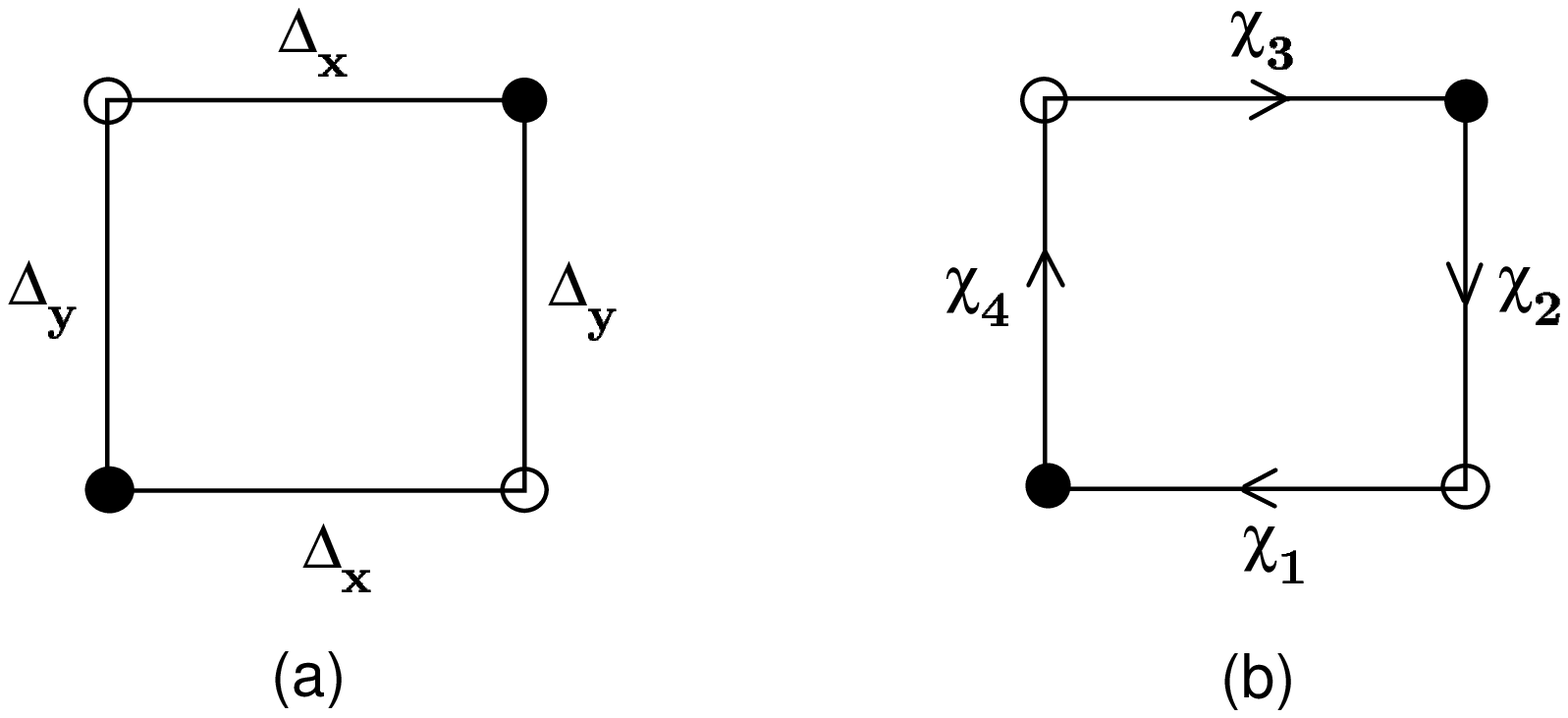}
\caption{\label{Fig1}}
\centerline{{\footnotesize 
The unit cell of the square lattice and the definition of }}
\centerline{{\footnotesize 
(a) the RVB and (b) the flux order parameters.}}
\end{center}
\end{figure}

\noindent
Due to the assumed translational
symmetry, it is convenient to introduce
a different notation for the fermionic fields living on odd and even
sites (empty and full dots in the figure), say $\psi$'s and $\phi$'s.
If we now go to momentum space -- with $k$ belonging to the Reduced
Brillouin Zone (RBZ) and $\omega_n$ denoting the fermionic
Matsubara frequencies -- the partition function in the static approximation
(i.e. the auxiliary fields being time independent) can be rewritten as:
\bea
\cz &=& \int [\cd \psi^*_{k\a,n}\, \cd \psi_{k\a,n} ]
 [\cd \phi^*_{k\a,n}\, \cd \phi_{k\a,n} ] \int [\cd \chi^*_{j}\,
\cd \chi_{j} ] [\cd \D_{j}^* \, \cd \D_{j} ]   \label{hs} \nonumber \\
 & \times &\exp\left\{ - \frac{2 M\b}{J}
\sum_{j=1}^4 |\chi_j|^2 -  \frac{4 M\b}{J} \sum_{j=x,y} |\D_j|^2 \right\}
\nonumber \\
& \times & \exp \left\{  \sum_{k\in RBZ} \sum_{\a} \sum_n \left[
\psi^*_{k\a,n} (i \o_n + \m \b) \psi_{k\a,n} +
\phi^*_{k\a,n} (i \o_n + \m \b) \phi_{k\a,n} \right] \right\} \nonumber \\
& \times & \exp \left\{ -\b \sum_{k\in RBZ} \sum_{\a} \sum_n \left[
\l_k^* \psi_{k\a,n}^* \phi_{k\a,n} + \l_k \phi_{k\a,n}^* \psi_{k\a,n}
 \right] \right\} \nonumber \\
& \times & \exp \left\{ -\frac{\b}{2} \sum_{k\in RBZ} \sum_{\a} 
\sum_n \Bigl[
\D_k^* \left( \psi_{k\uparrow,n} \phi_{-k\downarrow,-n}-
 \psi_{k\downarrow,n} \phi_{-k\uparrow,-n} \right) \right. 
\nonumber \\
& & \hspace{3cm}+ \hspace{3mm}\D_k^* \left( \phi_{k\uparrow,n} \psi_{-k\downarrow,-n}-
 \phi_{k\downarrow,n} \psi_{-k\uparrow,-n} \right) +  h.c. \Bigr] \Biggr\}
\label{FI}
\eea
where
\bea
 \l_k &=& (\chi_1+t\d) e^{ik_x} +  (\chi_2^*+t\d) e^{-ik_y} +
 (\chi_3+t\d) e^{-ik_x} +      (\chi_4^*+t\d) e^{ik_y}\hspace{1cm} \\
    \D_k &=& 2 ( \D_x \cos k_x + \D_y \cos k_y ) 
\eea
and $M$ is the number of lattice sites. 
The effective action in the functional integral (\ref{FI}) is now quadratic 
in the Grassmann fields $\psi,\phi$, and it is actually a sum of 
independent quadratic forms acting separately in the four-dimensional 
subspaces spanned by 
$(\psi_{k\a ,n},\phi_{k\a, n},\psi_{-k\overline{\a},-n}^*,\phi_{-k,\overline{\a},-n}^*)$, 
for $k$ in the RBZ, $\a = \uparrow, \downarrow$ and $\o_n = (2n+1) \pi$, 
where they are represented by the matrices: 
\be
i \o_n {\bf I}_{4\times 4} +\b \left( \begin{array}{cccc}
\m & \l_k^* & 0 & - \D_k \\
\l_k & \m & - \D_k & 0 \\
0 & - \D_k^* & -\m & -\l_k \\
-\D_k^* & 0 & -\l_k^* & -\m
\end{array} \right)       \label{mat}
\ee
Integrating then over the Grasmann fields, we arrive at the result:
\bea 
\cz&=&  \int [\cd \chi^*_{j}\cd \chi_{j}]  [\cd \D_{j}^* \cd \D_{j} ]
\exp \{ - \b \Omega \} \\
\Omega &=& \frac{2M}{J} \left[ \sum_{j=1}^4 |\chi_j|^2 +2
\sum_{j=x,y} |\D_j|^2 \right] + \cs_{eff} - \m M     \label{omega}
\eea
\bea
\cs_{eff}& =& -\frac{1}{\b} \sum_n \sum_{k\in RBZ} \sum_{j=1}^4 ( -i \o_n - \b
\log E_k^j )    \label{action} \\
&= &-\frac{1}{\b}   \sum_{k\in RBZ} \sum_{j=1}^4
\log \left( 1+ e^{\b E_k^j} \right) \; , \nonumber   
\eea
where $E^{j}$ ($j=1,2,3,4$) denote the eigenvalues of the matrix (\ref{mat})
multiplied by $\b$. Such matrix being $4\times 4$, it would be  
possible to give an analytic formula for the eigenvalues $E_k^j$ and
their dependence on the parameters $\mu,\l_k,\D_k$,
which is however very complicated and will not be presented here for the
general case.

From (\ref{omega}) one can then derive the self-consistency equation that must
hold for the parameters $\chi$'s and $\D$'s in the mean field approximation:
\bea
\frac{\partial \O}{\partial \chi_j} &=& 0 \;\;\; (j=1,2,3,4)  \label{chi}\\
\frac{\partial \O}{\partial \D_j} & =& 0 \;\;\; (j=x,y) \label{delta}
\eea
which have to be supplemented by the equation fixing the chemical potential:
\be
\frac{\partial \O}{\partial \m} = - M(1-\d) \label{mu}
\ee
We have numerically investigated the solutions to this set of equations that
also minimize the grand potential $\O$, finding that all solutions belong to
one of the following classes:

\begin{itemize}
\item[1)] \underline{Mixed-wave solutions}:
\be
\chi_1 = \chi_2 =\chi_3 =\chi_4 \equiv\chi \;\; \;
, \; \;\;
\D_x \equiv \D \; , \;    \D_y = e^{i\t} \D  \label{mix}\; .
\ee

\noindent
\item[2)] \underline{Dimer solutions}:
\be
\chi_j \mbox{ real, } \; |\chi_1| >> |\chi_2|=|\chi_3|=|\chi_4|
 \;\; \;\; \;\;
\D_x = \D_y = 0  \label{dimer}          \; .
\ee
\end{itemize}

Under this assumptions it is possible to give a nice formula for the
eigenvalues $E_k^j$ and hence for the grand potential.\\
The effective action for the mixed-wave solution reads:
\bea
S_{eff}& =& -\frac{2}{\b} \sum_{k\in RBZ} \left[ \log \cosh
\frac{\b E_k^+}{2} + \log \cosh \frac{\b E_k^-}{2} + 2\log2 \right]
\label{act} \\
E_k^\pm &=&   \sqrt{ (\m \pm |\l_k|)^2 + |\D_k|^2} \nonumber \\
\l_k &=& 2[(\chi+t\d) \cos k_x +  (\chi^*+t\d) \cos k_y ] \nonumber\\
\D_k &=& 2\D( \cos k_x + e^{i\t} \cos k_y ) \nonumber
\eea
For the dimer solution one finds:
\bea
S_{eff}& =& -\frac{2}{\b} \sum_{k\in RBZ} \left[ \log (1+ e^{
\b (\m -E_k)} ) + \log  (1+ e^{ \b (\m +E_k)} ) \right] 
\label{actd} \\
E_k &=&  |\l_k|  \nonumber\\
\l_k &=& (\chi_1+t\d) e^{i k_x} + (\chi_2^*+t\d) e^{-i k_y}+
(\chi_3+t\d) e^{-i k_x} + (\chi_4^*+t\d) e^{i k_y}  \nonumber
\eea

It is among these classes of solutions that one has to look for the
absolute minimum of the free energy $F= \O + \m M (1-\d)$ as the temperature
$T$ and the doping $\d$ vary. But, before presenting the phase diagram in
the $\d-T$ plane, we need to understand some more properties of such
solutions.
\begin{itemize}
\item[1)] The solution we call mixed-wave can actually represent different phases
already studied in literature, depending on the relative
values of the two independent complex parameters $\chi$ and $\D$:
\begin{itemize}
\item[a)] {\bf{Uniform phase}}: it corresponds to $\D=0$ and $\chi$ real.
As we will see this is the stable phase at high doping and high temperature.
\item[b)] {\bf{Flux phase}}: it corresponds again to $\D=0$ but now $\chi$ has a
nonvanishing imaginary part. If we write $\chi = |\chi| e^{i\phi}$ we
have that $4 \phi$ might be seen as a magnetic flux threading the elementary
square plaquette. One can check that, at fixed $T$, the value of the phase
$\phi$ increases while $\delta$ decreases, going from $\phi=0$ at high doping
(uniform phase) to $\phi=\pi/4$ at half filling ($\d=0$).
\item[c)] {\bf{RVB-d phase}}: it is obtained when $\chi=0$ and $\t=\pi$.
The actual value of the parameter $\D$ depends, at fixed T, on the doping
$\d$.
\item[d)] {\bf{RVB-mixed phase}}: it is given by $\chi=0$ and $\t=\pi/2$. Again the
value of the parameter $\D$ depends on the doping $\d$.
\item[e)] It is important to notice that, at half filling, the $\pi$-flux and
the RVB-mixed phases are equivalent via an $SU(2)$ transformation, and hence
are degenarate in energy and actually represent the same physical state.
It is only away from half filling that these two phases become independent.
Still, one expects them to remain very close in energy, at least for low
doping.
\end{itemize}

\item[2)] In the dimer solution, the relative values of the two independent order
parameters $\chi_1$ and $\chi_2$ change with $\d$, for fixed $T$.
In particular, at any temperature, $\chi_2=0$ at half filling, so that only
one link variable per plaquette is nonzero, giving rise to the so called
staggered dimer configuration.    
\end{itemize}

As for solutions of type (1), our analysis has shown that none of the above
mentioned ``pure solutions'' is the actual absolute minimum of the free energy.
At not too small values of $\d$, the latter is indeed minimized by a
phase in which both the $\chi$ and the $\D$ parameters are different from
zero. Precisely we obtain a sort of mixture between the uniform and the
RVB-d phases, which we would like to call the {\it mixed phase}:%
\footnote[7]{This phase is simply referred to as the ``dimer phase'' 
in \cite{Z,KL}}
\be
\chi \mbox{ real \ \ and \ \ } \t = \pi \; .\label{mixed}
\ee
The values of $|\chi|$ and $|\D|$ change with $T$ and $\d$. Above a critical
temperature $T_c = J/8$ we find $\D=0$ and hence recover the uniform
solution. At any $T<T_c$, there is a critical doping $\d_c$ above which
the stable solution  is again the uniform phase. Below $\d_c$, the parameter
$\D$ increases continuously (second order transition) while $\d$ decreases,
in such a way that for $\d=0$ one also has $|\chi| = |\D|$. It is interesting
to notice that the solution we find at
half filling, where the $SU(2)$ symmetry is restored, is indeed equivalent
via a  gauge transformation to the RVB-mixed phase, or to the
$\pi$-flux phase. This means that the introduction of the kinetic term
breaks the $SU(2)$-degeneracy in favour of a state where all the $\chi$'s
are equal and real, while the $\D$'s differ just for a phase of $\pi$.

At low dopings, it is the dimer phase that becomes stable. The transition
from the dimer to the mixed phase is a first order transition.

The phase diagram is shown in Fig. \ref{Fig2} for $t/J = 1$. For
different values of  $t/J$ one obtains diagrams that look qualitatively the
same: what changes slightly is only the position of the transition lines.
\begin{figure}[t]
\begin{center}
\epsfxsize=7cm \epsffile{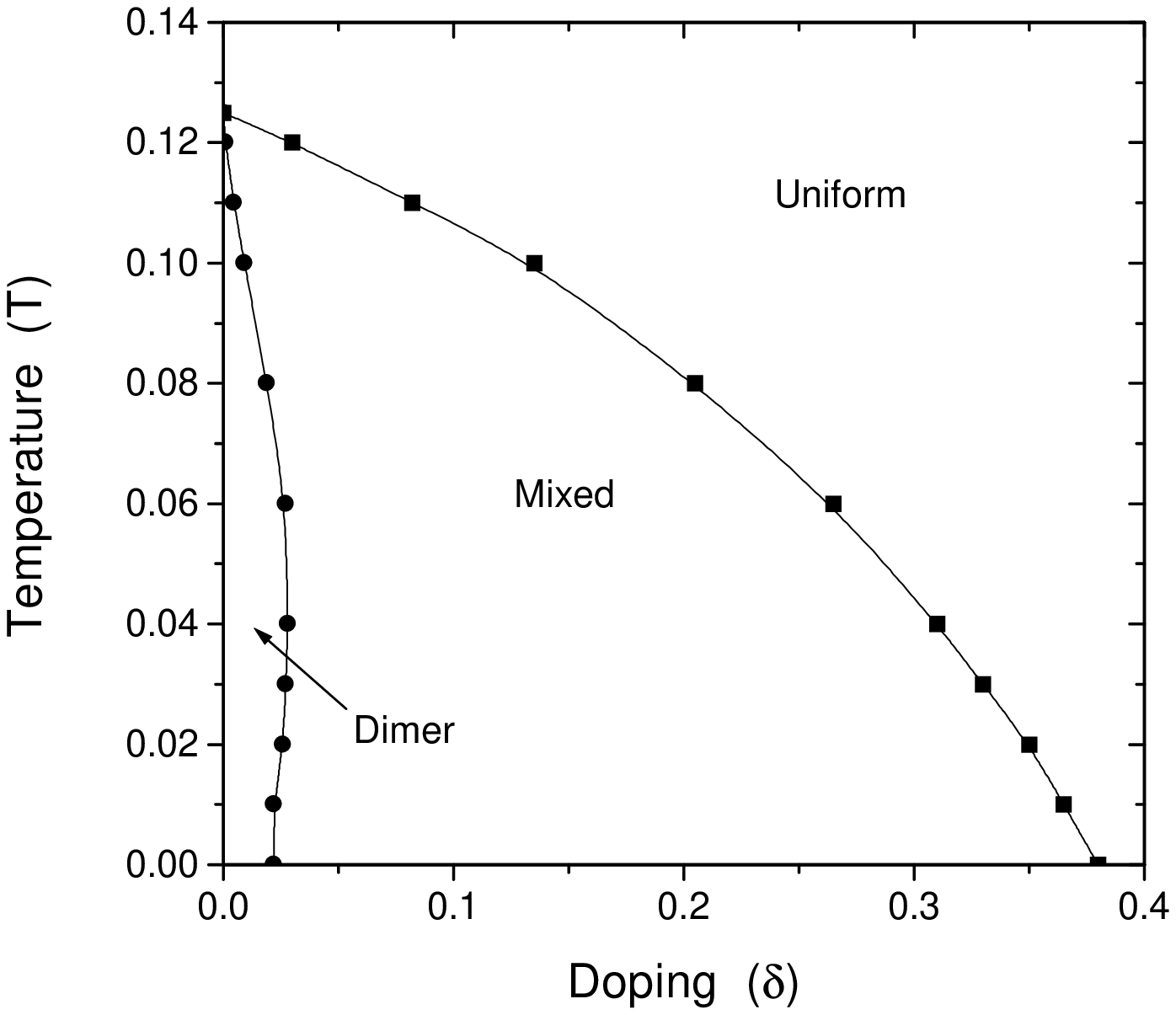}
\caption{\label{Fig2}}
\centerline{{\footnotesize 
The mean-field phase diagram of the $t$-$J$ model.}}
\end{center}
\end{figure}

We have also investigated the stability of the mean field solutions
around the lines of transition,
by looking at the behaviour of the chemical potential
$\mu$ as a function of doping $\d$, since an instability towards phase
separation is signaled by $\paf{\mu}{\d}>0$ (see the first paper in
\cite{EPR} for a discussion).

While no instability develops at the mixed/uniform line of
transition, we find that the coexistence of
the dimer and the mixed phases is favoured at all temperatures $0 < T <
T_c$ around the original line of transition. Such effect is pretty strong
especially at low temperatures, so that the pure dimer phase survives
only very close to half filling. This is shown in Fig. \ref{Fig3}.
\begin{figure}[t]
\begin{center}
\hspace{-0.1cm}
\epsfxsize=7cm \epsffile{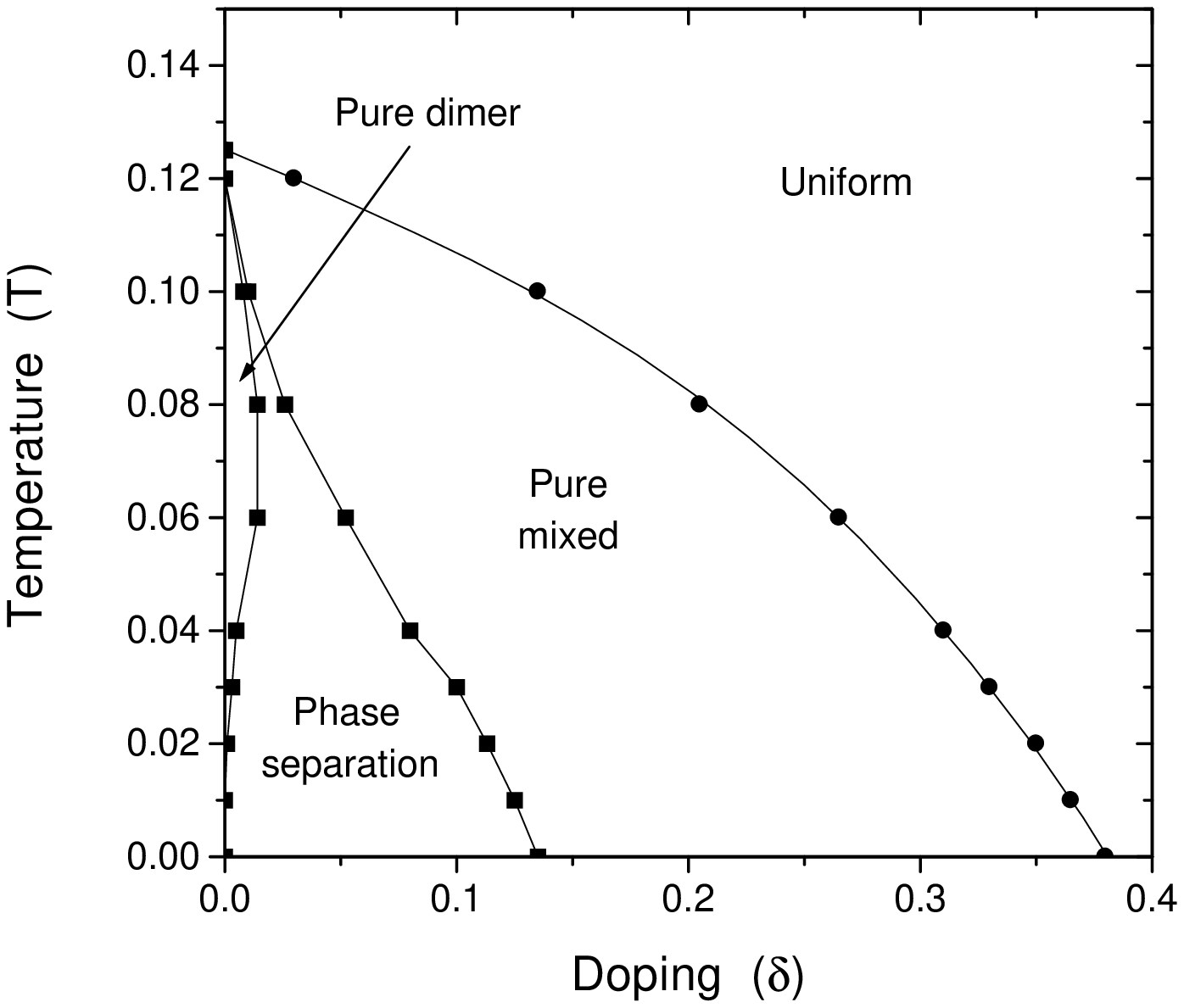}
\caption{\label{Fig3}}
\centerline{{\footnotesize 
The phase diagram of the $t$-$J$ model, including the }}
\centerline{{\footnotesize 
occurrence of phase separation.}}
\end{center}
\end{figure}

We would like now to compare these results with our previous work
\cite{EPR}. Even if the starting Hamiltonian and the formalism we adopted are
the same, and despite the fact that in the present paper we have enlarged
the class of allowed mean field solutions, we find here a much simpler phase
diagram. In particular we notice that the flux phase is nolonger present,
at any value of the doping and temperature, being substituted by the mixed
phase (\ref{mixed}). This is in agreement to and extends the findings of
Zhang at zero temperature \cite{Z}. Also, our mean field Hamiltonian
and self-consistency equations coincide, but for some normalization
coefficients, with the one given in \cite{KL} and obtained through an
auxiliary-boson approach, originally formulated by Kotliar and Ruckenstein
\cite{KR} for the Hubbard model, in which an empty site is described by a
bosonic ceation operator $b_i^\dagger$, whose density gets to be proportional
to the doping in the meand field approximation:
$\langle b_i^\dagger b_j\rangle = \d${\footnote[3]{Parenthetically, we notice
that this explains also why the hopping Hamiltonian, which is written as
$t\sum_{(ij)} \sum_{\a} c^\dagger_{i\a} c_{j\a} b_j^\dagger b_i$ in this
approach, becomes $t\d \sum_{(ij)} \sum_{\a} c^\dagger_{i\a} c_{j\a}$
in the mean field approximation as we wrote in (\ref{tj}).}}.

On the contrary, the authors of ref. \cite{CZ} find some wide
region where the flux phase is stable and a small area of the phase diagram
where there is a coexistence of the flux and the superconducting phases.
They represent the $t$-$J$ model via the so-called Hubbard $X$-operators,
which allow for the enforcemeent of the below-half filling constraint,
and perform a $1/N$ espansion to extrapolate then to the case $N=2$.

Our phase diagram is also substantially different from the one
presented by Ubbens and Lee in \cite{UL} and obtained within
a slave boson approach that differs from the one mentioned above \cite{KL} 
by the way the hopping term is treated. The free energy
and hence the self-consistency equations found in \cite{UL} have an additional
contribution coming from all the (infinite) bosonic states. As a result,
the mean-field phase diagram shows a finite region, close to half filling,
where the flux solution (either $\pi$ or staggered) 
is favoured against all other
phases. A similar result is obtained, at zero temperature,
also by Sheng {\it et al.} in \cite{SSY}. They work
again within the slave boson approach by introducing now an additional
statistical gauge field that describes the hard-core nature of holes. In
this case the free energy gets minimized, at low dopings, by a staggered
flux phase combined with a nonzero statistical field yielding a uniform
flux per plaquette. 

As we have also noticed in the second paper of ref.
\cite{EPR}, our calculations (\`a la Anderson) and the slave
boson techniques seem to lead to very different results, in particular
close to half filling.
We are investigating why this discrepancy occurs. Even if we do
not have a final answer to this question, we believe that the problem has
to do with the way the below-half filling constraint is treated
within the different approaches. Indeed, at the mean field level
such constraint is respected only in some average sense and differences
in the way this is implemented can result in different phase diagrams,
especially close to half filling when the high density of electrons
enhances the probability of including configurations with double occupied
sites.

Finally, we want to comment on the possibility that, at a fixed temperature,
phase separation might disappear when $J/t$ is lowered below a critical
value $J_c/t$. In the second paper of ref. \cite{EPR}, we studied how this
happens when considering non-magnetic phases only, showing that $J_c/t$ varies
linearly with $T$ according to $J_c/t\approx 8 T$. This results is confirmed
in our present calculations, where we have also included RVB states.
There is a difference however: now phase separation occurs as soon as the
temperature is lower than the critical temperature $T_c =  J/8$, whereas
in our previous paper at temperatures close to the critical one only
pure phases were present.

\vskip 0.5cm
{\bf Acknowledgments}

The authors would like to thank P. Pieri for the many 
helpful discussions. 

\end{document}